# Superior visible photoelectric response with Au/Cu$_2$NiSnS$_4$ core-shell nanocrystals

Anima Ghosh[1,2 #], Shyam Narayan Singh Yadav[3#], Ming-Hsiu Tsai[4], Abhishek Dubey[3], *Shangjr Gwo[5,6],* Chih-Ting Lin[4], and Ta- Jen Yen[3*]

[#]Equal contribution

[1]Department of Physics, School of Sciences, SR University, Warangal-506371, India

[2]Institute of Atomic and Molecular Science, Academia Sinica, Taiwan-101607

[3]Department of Materials Science & Engineering, National Tsing Hua University, Hsinchu, Taiwan-30013

[4]Graduate Institute of Electronics Engineering, National Taiwan University, Taipei 1061, Taiwan, ROC

[5]Department of Physics, National Tsing Hua University, Hsinchu 300, Taiwan R.O.C.

[6]Research Centre for Applied Science, Academia Sinica, Taipei 115-29, Taiwan R.O.C.

*corresponding author*

**Abstract**

The incorporation of plasmonic metal nanostructures into semiconducting chalcogenides, in the form of core-shell structures, represents a promising approach to boosting the performance of photodetectors. In this study, we combined Au nanoparticles with newly developed copper-based chalcogenides $Cu_2NiSnS_4$ (Au/CNTS), to achieve an ultrahigh optoelectronic response in the visible regime. The high-quality Au/CNTS core-shell structure was synthesized by developing a unique colloidal hot-injection method, which allowed excellent control over sizes, shapes, and elemental compositions. The fabricated Au/CNTS hybrid core-shell structure exhibited enhanced optical absorption, carrier extraction efficiency, and improved photo-sensing performance, owing to the plasmonic-induced resonance energy transfer effect of the Au core. This effect led to a significant increase in carrier density between the Au core and CNTS shell, resulting in measured responsivity of $1.2 \times 10^3$ AW$^{-1}$, specific detectivity of $6.2 \times 10^{11}$ Jones, and external quantum efficiency (EQE) of $3.8 \times 10^5$ %, respectively, at an illumination power density of 318.5 µWcm$^{-2}$. These values outperformed a CNTS-based gate-free visible photodetector.

**Introduction**

The development of earth-abundant chalcogenide materials has various applications such as spectroscopy, optoelectronics, and photovoltaics.[1-3] Instead of conventionally expensive and toxic Cd- and Pb-based chalcogenides compounds, recent progress on copper-based chalcogenides such as $Cu_2X$(X=Zn, Fe, Co, Ni, Mn)$SnS_4$ (i.e., CXTS), further draws extensive attention owing to their excellent optoelectronic properties, including *p*-type conductivity, direct band gaps of ~1.2-1.5 eV.[4-5] Moreover, a high absorption coefficient in the visible range,[6-10] makes them attractive as p-type photo-absorbing layers for optoelectronic devices.[3, 11-18] Among various CXTS compounds, $Cu_2NiSnS_4$ (CNTS), was found to show an optical absorption coefficient ($\approx 10^6$ cm$^{-1}$) and very low conduction band offset ($-0.12$ eV) [4-5, 7, 19-21]. To further employ CNTS for practical optoelectronic applications, a demanded challenge is to further boost its optical absorption.[4, 14] As a consequence, there are various methods to enhance CNTS's optical absorption and optoelectronic responses, such as the hybrid of inorganic core-shell nanostructures [22-24] and the introduction of plasmonic resonances.[25] Especially, substantial effort has been focused on synthesizing metal-semiconductor hybrid structures that offer unique interfacial electronic behaviors, superior band alignments, and excellent performance for optoelectronic devices.[15, 19, 21, 26] The metal/semiconductor hybrid nanostructures have opened up a new path by incorporating plasmonic metallic nanoparticles (NPs) to passivate surface states with enhanced optical properties.[27-29] Hence, these incorporations of plasmonic NPs in the semiconductor NCs can improve charge extraction and collection,[30] which is appropriate for achieving higher optoelectronic performance.[31-35]

In this work, we synthesized Au/CNTS core-shell nanostructures by developing a unique colloidal hot injection method, which allows tuning the chemical compositions, crystal structures, particle sizes and surface morphology to significantly enhance the optical properties.[36-37] An immense list of metal salt precursors, together with chalcogen sources, started nucleation in solution with a relatively low temperature, which amenably stabilized the Au/CNTS core-shell nanocrystals (NCs). The Au core enhanced the optical absorption of CNTS by 35% based on the plasmonic-induced resonance energy transfer (PIRET) effect, evidenced by the UV-visible absorption measurement and the finite difference time domain (FDTD) simulation. Such enhancement suggests that these photoactive core-shell NCs are excellent photo absorbents in the visible region for optoelectronic devices. Furthermore, we realized the first-ever Au/CNTS core-shell-based photodetector, demonstrating high responsivity of $1.2 \times 10^3$ AW$^{-1}$, specific detectivity of $6.2 \times 10^{11}$ Jones, and external quantum efficiency (EQE) of $3.8 \times 10^5$ % and considerable response/recovery times of 2.58/11.14 S, along with excellent operational reliability. Our results demonstrate the potential of plasmonic core-shell nanostructures for the development of visible photo-sensing devices for optoelectronic applications.

**Results and Discussions**

Our designed core-shell nanostructure and a graphene-based hybrid photodetector is illustrated in **Figure 1**a. **Figure 1**b shows the enlarged image of the selected area of **Figure 1**a. In this hybrid photodetector, the core-shell nanostructure consists of the Au core as a plasmonic NP, enabling localized surface plasmon resonance (LSPR,) and the shell CNTS nanostructure works as a photoactive semiconductor. Additionally, owing to high conductivity,[38] ultra-high charge carrier mobility ($10^5$ cm$^2$ V$^{-1}$ s$^{-1}$),[39] and CMOS compatibility, graphene works as a channel layer to transport excited charge carriers under the applied bias voltage across the two electrodes. We synthesized CNTS and Au/CNTS NCs at a temperature of 200°C using the colloidal hot-injection method[10, 26] in an inert atmosphere (refer to Experimental Section for details). The optimized NCs were obtained by varying the growth time, thiol injection amount, time and temperature, and other related parameters. We observed that at a temperature of less than 200°C, the incomplete nucleation and growth process occurs due to a random shape and size, resulting in the impure phase of the NCs.

Next, the structural and elemental composition of these as synthesized NCs were scrutinized using the powder X-ray diffraction (XRD) spectroscopy, high-resolution transmission electron spectroscopy (HRTEM), energy-dispersive X-ray spectroscopy (EDXS), and X-ray photoelectron spectroscopy (XPS), respectively. The XRD spectra of CNTS and Au/CNTS NCs are shown in **Figure 2**a. The XRD spectra peaks of CNTS NCs at 28.43°, 33.06°, 47.49°, 50.87°, and 56.15° represent the (111), (200), (220), (222), (311) planes, respectively. The XRD pattern confirms the pure cubic crystal structure of CNTS [JCPDS data no. 26-0552].[4, 40] Also, the XRD spectra (blue color line) for Au/CNTS core-shell NCs indicate the cubic phase of CNTS, and the cubic lattice of the Au core is evident from the (111) reflection of the face-centered cubic lattice. There are no prominent peaks caused by any impurities. The TEM images of CNTS and Au/CNTS NCs are shown in **Figure 2**b and **Figure 2**c, respectively. **Figure 2**b revealing the average particle size of CNTS NCs is ~13.5±2.4 nm. **Figure 2**c exhibits that the resulting NP of Au incorporated CNTS is the core-shell structure of 0.05 M Au with CNTS, not the mixture or ad-mixture of individual Au nanoparticles and CNTS NCs.

The selected area electron diffraction (SAED) patterns are displayed in **Figure 2**d, which indicates that CNTS nanocrystals are polycrystalline and pure cubic crystal structure. The concentric rings of this SAED pattern agree well with the XRD analysis. The HRTEM of CNTS NCs is shown in **Figure 2**e, demonstrating that the CNTS has an interplanar distance of 0.31 nm, which indicates the (111) plane of a cubic crystal. The HRTEM of Au/CNTS is shown in **Figure**

**2**f, demonstrating an interplanar distance of 0.21 nm corresponding to the (111) plane of cubic Au NP in the core. The STEM image and elemental mapping of CNTS NCs are shown in **Figure S**1 and **Figure S2** (Supporting Information), displaying their distribution and validating the pristine CNTS and core/shell structure of Au NPs and CNTS NCs. The chemical compositions of CNTS and Au/CNTS NCs are verified from the energy-dispersive X-ray spectroscopy (EDXS) analysis. The EDXS study (**Figures S3** and **Figure S4**, Supporting Information) reveals that the atomic Cu: Ni: Sn: S ratio is close to the stoichiometry of $Cu_2NiSnS_4$. Further, EDX analysis confirmed that the Au/CNTS core-shell NCs containing ~ 5.24 wt% Au (**Figure S4**, Supporting Information) were prepared from 0.05 mmol of $HAuCl_4·3H_2O$.

The X-ray photoelectron spectroscopy (XPS) measurements were performed (**Figure S5**, Supporting Information) on the surface of the NCs, to determine the oxidation states of the constituent elements. The XPS gives a suitable complement to EDX for chemical composition analysis. The Cu *2p* core level spectrum (**Figure S5a**, Supporting Information) shows peaks at 933.25 eV ($2p_{3/2}$) and 953.15 eV ($2p_{1/2}$) with a peak separation of 19.9 eV, confirming the $Cu^+$ state. The Ni (II) peaks at 858.08 eV ($2p_{3/2}$) and 875.24 eV ($2p_{1/2}$) correspond to the $Ni^{2+}$ state, with the splitting value of 17.16 eV (**Figure S5b**, Supporting Information). The doublet peaks of Sn (IV) are separate and located at 494.05 eV ($3d_{3/2}$) and 485.05 eV ($3d_{5/2}$), shown in **Figure S5c** (Supporting Information). **Figure S5d** confirms the sulfur spectrum for the $S^{2-}$ level from the $2p_{3/2}$ and $2p_{1/2}$ peaks with a doublet separation of ~1.09 eV.

Furthermore, we characterized the CVD-grown graphene (for synthesis details, refer to Experimental Section) using the Raman spectroscopy, and corresponding spectra are shown in **Figure S6** (Supporting Information), demonstrating three feature peaks D, G, and 2D bands at 1340 cm$^{-1}$, 1580 cm$^{-1}$, and 2690 cm$^{-1}$, respectively. The intensity ratio ($I_{2D}/I_G$) of the 2D and G peaks was found to be 2.34. Both the expected peak position and intensity ratio confirmed that the transferred graphene layer at the top of $SiO_2/p^+$-Si is a single-layer graphene.[41-42] Next, we examined the absorption of both CNTS and Au/CNTS hybrid nanostructure, to reveal the enhanced absorption efficiency by introducing Au NPs using plasmonic effect. The corresponding UV-Visible spectra are shown in **Figure 3**a. It is important to note that the pristine CNTS structure has a lower optical absorption efficiency in contrast to the Au/CNTS nanostructure. CNTS and Au/CNTS NCs both demonstrated a lower absorption efficiency at longer wavelengths and become higher toward shorter wavelengths.

To study the enhancement in optical absorption due to the Au core in the CNTS shell, we conducted a numerical simulation by using Lumerical® software finite difference time domain (FDTD) (refer to Experiment Section for the simulation details). The scattering and absorption efficiency were calculated for both pristine CNTS and Au/CNTS hybrid NCs, as shown in **Figure S7** (Supporting Information). The calculated extinction efficiency is depicted in **Figure 3**b and is in accordance with measured UV-Vis spectra for both CNTS and Au/CNTS. The Au NP in core dramatically enhances the ~~poor~~ absorption of the pristine CNTS NCs, in particular around its resonance wavelength because of the incident field localization caused by the induced LSPR. Additionally, the increased optical path length of incident photons increases light-matter interaction within the semiconductor, which increases optical absorption.[28, 43] The electric field distribution was plotted at the highest absorption efficiencies wavelength of 405 nm and is depicted in **Figure 3**c, illustrating that the electric field is highly confined around the NCs. This highly confined field induces hot electron generation and PIRET effect takes place from the surface of the metal to the semiconducting CNTS NCs, simultaneously. Moreover, we draw a Tauc plot to calculate the band gap of synthesized pristine CNTS,[44] depicted in **Figure 3**d. The band gap is calculated by intercepting the slope along the x-axis and is found to be 1.54 eV, matching well with the reported literature.[4]

To explain the charge carrier generation and the transfer mechanism in core-shell Au/CNTS and graphene hybrid system, the schematics of the energy band diagram with an applied biased are illustrated in **Figure** 4. Here, the graphene has a fermi energy of −4.6 eV, [45-46] and Au has a work function of −5.1 eV. When metal (Au) and semiconductor (CNTS) are brought into contact, they form a Schottky barrier $Φ_B$. At the Au NPs and CNTS interface, Au NPs plays an important role to enhance the population of charge carriers in the CNTS via nonradiative plasmonic decay as listed. (1) when light is incident on Au/CNTS system, plasmon-induced hot electrons are generated at the surface of Au and transfer to CNTS, (2) plasmonic-induced charge transfer transition (PICTT) directly generates charge carriers in the conduction band of CNTS, and (3) PIRET effect increased the population density of photogenerated excitons in CNTS NCs.[29] These photogenerated charge carriers reach their corresponding energy level when $V_{ds}$ = 0 V and there will be no net current flow as depicted in **Figure 4**a. In contrast, under the applied bias across the electrodes, the charge carriers drift toward the opposite polarity, and there will be a net current flow through the device, as shown in Figure 4(b).

Next, we probed the synthesized Au/CNTS NCs optoelectronic characteristics, and demonstrated its excellent performance as a visible photodetector. First, we fabricate two separate photodetectors using the bottom-up fabrication technique, as shown in **Figure S9** (Supplementary Information), using monolayer graphene as a conducting channel layer with pristine CNTS and Au/CNTS hybrid NCs as photo absorbing layer. The cross-section view of the designed photodetector is shown in **Figure 5**a. The photocurrent was probed using applied bias voltage ($V_{ds}$) across two electrodes (drain and source).[47] The dark/light current ($I_{dark}/I_{light}$) was measured without/with light illumination. The dark current is found to be of the order of µA, due to thermally excited excitons at the surface of the graphene caused by the gapless band characteristic of graphene.[48] The measured photocurrent ($I_{ph} = I_{light} - I_{dark}$) and the transient response for the CNTS and Au/CNTSan at an incident wavelength of 405 nm at a fixed illumination power of 860 µW are shown in **Figure 5**b and **Figure 5**c. A significant photocurrent enhancement has been observed for the Au/CNTS core-shell NCs-based photodetector in contrast to the pristine CNTS NCs-based photodetector.

The enhancement in the photocurrent can be understood by the following mechanisms: There is an enhanced scattering efficiency due to the introduction of the Au core in the CNTS NCs which increases the optical path length, resulting in high optical absorption. In plasmonic metal/semiconductor photodetectors, the close contact between metal and semiconductor enables efficient charge transfer by strong LSPR of the noble metal, effectively reducing the charge carrier recombination dipole-dipole coupling of excitons and plasmon.[49] These aforementioned mechanisms will help CNTS to enhance the population of excited excitons. These excited excitons disassociated under an applied field across the two electrodes (drain and source) and drift toward opposite polarities resulting in the net photocurrents. To investigate the photo response characteristic of the photodetector, the key parameters, e.g., photoresponsivity ($R_\lambda$), specific detectivity ($D^*$), and external quantum efficiency (*EQE*), were calculated by the following equation (1-4),[50-51]

$$R = \frac{I_{photo}}{P};  \quad (1)$$

$$D^* = \frac{R_\lambda}{\sqrt{\frac{2qI_{dark}}{A}}}; \quad (2)$$

$$D^* = \frac{\sqrt{AB}}{NEP}; \tag{3}$$

$$EQE = \frac{(\hbar c R_\lambda)}{q\lambda}. \tag{4}$$

Here $I_{photo}$, $P$, $R_\lambda$, $q$, $I_{dark}$, and $A$ denote the photocurrent ($I_{photo} = I_{light} - I_{dark}$), illumination power, responsivity at a wavelength of $\lambda$, electric charge, dark current, and active area of the device, respectively. Next, $B$, $NEP$, $h$, $c$, and $\lambda$ denote the bandwidth, noise equivalent power, Planck constant, speed of the light, and wavelength, respectively.

$$NEP = \frac{I_N}{R}; \tag{5}$$

Here, $I_N$ is the noise current, and is defined as $I^2_N = 2qI_{dark}B$.

The transient photocurrent response for both photodetectors was measured by applying constant bias voltage $V_{ds}$ of 2 V, as shown in **Figure 5**c. The results show that when the light was allowed to expose, the current increased with time, and it saturated after reaching its maximum due to the saturation in optical absorption and recombination of excess excitons that could not be collected through electrodes before they were annihilated. The calculated photoresponsivity ($R_\lambda$) for pristine CNTS and Au/CNTS hybrid NCs-based photodetector with the corresponding illumination power densities are shown in **Figure 5**d. Here, we observed that the responsivity for Au/CNTS hybrid NCs was enhanced in contrast to pristine CNTS NCs and decreased with higher illumination power density. This decrease in responsivity with higher power is caused by saturation in optical absorption, the screening of the field by photoexcited careers, and enhanced career scattering rate.[52-54] To examine the photo response speed, the rise time ($\tau_r$) and fall time ($\tau_f$) are calculated. $\tau_r$ is defined by the time it takes for the maximum photocurrent to reach from 10% to 90%, whereas $\tau_f$ is the duration when the maximum photocurrent reduces from 90% to 10%.[55] The calculated rise/fall time for CNTS and Au/CNTS-based photodetectors was found to be 2.6/3.4 S and 3.4/13.14 S as shown in **Figure S10.** The rise and fall time mainly depend on the defect and trap densities of the bandgap region of semiconductors. [56-57] The slow response time is due to the photogating effect due to the presence of a trap state in the fabricated photodetector. The response speed could be further improved by engineering the design of the photodetector.[58]

Furthermore, to explore the broadband response in the visible range of our fabricated photodetector, we conducted photocurrent measurements with three wavelengths 405, 532, and 632 nm of laser illumination under the visible spectrum. The time-dependent photocurrent with various illuminated power densities at a constantly applied voltage of 2V is shown in **Figure 6**a-c. The calculated photoresponsivity and detectivity for 405 nm, 532 nm, and 632 nm wavelengths with illumination power density are shown in **Figure 6**d-f. The maximum photoresponsivity and detectivity were achieved at 405 nm wavelengths, with the lowest incident illumination power caused by the highest optical absorption at the shorter wavelength. The similar characteristics were observed for *EQE (***Figure S11**, Supporting Information*)*. It was found that the CNTS and Au/CNTS-based hybrid photodetectors are sensitive to all wavelengths in the visible spectrum.

The photocurrent spectral response of the photodetector with wavelength at fixed illumination power of 860 µW is shown in **Figure 7**a. The corresponding calculated responsivity and detectivity with wavelength at the fixed illumination power are shown in **Figure 7**b. Our designed photodetector has a broadband response for visible light with the best performance at 405 nm wavelength. The better optoelectronic performance (responsivity, detectivity, and EQE) at shorter wavelengths can be understood by UV-Vis absorption spectra and extinction efficiency shown in **Figure 3**a and **Figure 3**c, respectively. The results indicate that maximum photoelectric responses are achieved using an excitation laser of 405nm in contrast to longer wavelengths that use 532 and 632nm excitation lasers, which are off-resonance with the LSPR excitation wavelength. Finally, the achieved optoelectronic characteristics were compared with other reported work with a similar device configuration and are shown in Table S1 (Supporting Information) confirming that our Au/CNTS based photodetector achieved superior optoelectronic performance among other CNTS based Photodetectors.

**Conclusions**

To summarize, we synthesized a noble CNTS and Au/CNTS shell hybrid nanocrystal (NCs). The XRD and HRTEM results show that our synthesized CNTS and Au possess a pure cubic and face-centered cubic crystal structure with an interplanar distance of 0.31nm and 0.21 nm, respectively. The XPS results indicated that synthesized nanocrystals constitute all the elements with reasonable oxidation states. Furthermore, the optical studies of CNTS and Au/CNTS NCs indicate that Au/CNTS core-shell hybrid NCs possess significantly higher optical absorption than pristine CNTS NCs. Using these NCs, we fabricated a superior hybrid broadband photodetector and investigated its optoelectronic performance. Using Au/CNTS hybrid photodetector an ultrahigh responsivity of $1.2289 \times 10^3$ AW$^{-1}$, specific directivity ($\sim 6.18 \times 10^{11}$ Jones), and large EQE of $3.76 \times 10^5$ % at an incident wavelength of 405 nm was achieved. Notably, the enhanced optoelectronic performance was achieved at shorter wavelength because of the enhanced optical absorption induced by the LSPR effect of the Au core in Au/CNTS. The induced LSPR confines the incident field in the vicinity of the Au core and enhances the population of excited charge carriers significantly, resulting in enhanced photocurrent. Further, contact optimization, interfacial engineering, and optimizing device parameters could achieve a high photo-detecting device. This Au/CNTS core-shell NCs-based photodetector has many potential optoelectronic applications.

**Experimental Methods**

*Materials:* Copper (II) acetylacetonate (Cu(acac)$_2$, ≥99.9%), Nickel (II) acetylacetonate hydrate Ni(acac)$_2$, 99.99%), Tin (II) chloride (≥99.99% trace metals basis), Gold (III) chloride (≥99.99%), n-dodecanethiol (n-DDT, ≥98%), tert-dodecanethiol (t-DDT, 98.5%) and oleylamine (OAm, ≥98%), were purchased from Sigma-Aldrich. In addition, hexane (fraction from petroleum), Acetone, and Ethanol (anhydrous, Merck) were used without further purification.

*Synthesis of Cu$_2$NiSnS$_4$ (CNTS) NCs and Au incorporated Cu$_2$NiSnS$_4$ (Au/CNTS) core-shell NCs:* The colloidal hot-injection method was used to synthesize quaternary composition chalcogenide CNTS NCs and the Au/CNTS core-shell. In a typical synthesis of CNTS, 1 mmol Cu(acac)$_2$, 0.5 mmol Ni(acac)$_2$, and 0.5 mmol SnCl$_4$·5H$_2$O were mixed with 10 mL oleylamine (OAm) in a three-neck flask and stirred under vacuum for 60 min at room temperature. Then the system was back-filed with nitrogen, and the solution was subsequently heated up to 130°C. The reaction was followed by a quick injection of a mixture of 1-DDT and t-DDT (1:1 ratio) under a nitrogen atmosphere with continuous stirring. Subsequently, the solution was heated up to 200°C and maintained for the optimum time for completing the reaction. After the reaction, the mixture was cooled down to room temperature, and the final products were obtained via centrifugation with a mixture of hexane and ethanol. The as-synthesized nanocrystals (NCs) were readily dispersible in nonpolar solvents, e.g., octane, hexane, etc.

A similar procedure was adopted for the synthesis of Au/CNTS NCs using an Au nanoparticles (NPs) solution. The 0.02 mmol HAuCl$_4$·3H$_2$O was dissolved separately in oleylamine and heated to 120°C with a continuous stirrer for 10 M for Au NPs growth. Finally, Au NPs solutions were rapidly injected into the Cu-Ni-Sn precursor in the flask. For photodetection device fabrication, the as-synthesized CNTS, and Au/CNTS core-shell NCs were dispersed in n-hexane with a 50mg/ml concentration (optimized).

*Graphene Synthesis and Transfer:* Chemical Vapor Deposition (CVD) was utilized to synthesize a high-quality single layer with a large area of graphene (SLG) film on copper foil (99.8%, 25 μm thick) in a tabular quartz furnace. During the growth of SLG, the tube temperature was raised to 1000 °C in 80 min under continuous 110cmm $H_2$ gas flow. While the tube was at 1000 °C, $CH_4$ gas of 11 sccm flowed into the quartz tube and continued for 60 min for graphene to grow. The SLG can be obtained on the Cu substrate after the furnace cooled down under ambient $H_2$ gas and $CH_4$ gas. Then the SLG was transferred from Cu foil to a 300 nm $SiO_2$/Si substrate using the method of electrochemical delimitation.[47] CVD-grown single-layer graphene was transferred by wet transferred method on pre-cleaned $SiO_2$/Si substrate with Acetone, IPA, and Di water shown in Figure S8 (step II). The as-synthesized transferred graphene at $SiO_2$/Si substrate was characterized by Raman spectroscopy.

*Material Characterizations:*

XRD spectra were recorded with a Rigaku X-ray diffractometer (Cu Kα irradiation, λ = 1.541 Å). TEM, HRTEM, and HAADF-STEM images were taken from JEOL, JEM-2100F. For TEM analysis, the NCs dispersed in hexane and were drop cast on 300 mesh Ni grids. UV-vis absorption spectra were recorded by using Jasco V-670 spectrophotometer. The core-level XPS spectra of the NCs were carried out using Al Kα radiation (1486.6 eV) with base pressure $1.2 \times 10^{-8}$ Torr in photoelectron spectrometer PHI 5000 Versa Probe II, FEI Inc and was calibrated against C *1s* core level spectra at 284.8 eV at a base pressure of $5 \times 10^{-11}$ Torr. Raman spectra of graphene were collected using micro-Raman spectroscopy (HORIBA, LabRAM, HR800) with 532 nm solid-state laser excitation.

*FDTD Simulation*: Scattering spectra and field distributions were calculated using the three-dimensional finite difference time domain (FDTD) method with a commercial Lumerical® software package. A perfectly matched layer (PML) boundary condition was applied with a mesh size of 1 nm over all the fields including the Au core and CNTS shell nanoparticle. Total field scattered field source was used with the normal incident over 400 to 1000 nm. The refractive index of Au (gold) [59] and of CNTS was used from reported data in reference. [s4]

*Optoelectronic measurement:* The photo response of fabricated devices was characterized by a probe station system (Keithley, 4200 SCS) using a light source of visible region 405, 532, and 632 nm laser to excite the photoactive layer.

**Acknowledgments**

A.G. acknowledges DST-SERB (New Delhi) for financial support during the work. T.J. Yen acknowledges the financially supported by the "High Entropy Materials Center" from The Featured Areas Research Center Program within the framework of the Higher Education Sprout Project by the Ministry of Education (MOE) and the Project NSTC 111-2634-F-007-008 - by National Science and Technology Council (NSTC) in Taiwan.

**Authors Contribution**

A.G., S.N.S. Yadav and T.J. Yen conceived the project. T.J. Yen directed the project. A.G. synthesized, optimized the nanocrystals, characterized and analyzed the data of CNTS and Au/CNTS materials. S.N.S. Yadav, performed FDTD simulation. M.H. Tsai and C.T. Lin synthesized the graphene and characterized it. S.N.S. Yadav designed and fabricated the photodetection devices. S.N.S. Yadav and A. Dubey characterize the optoelectronic characteristics. S.N.S. Yadav processed the optical and optoelectronic data of the device. A.G.,

S.N.S. Yadav, and T.J. Yen participated in the preparation of the manuscript and commented on its content.

**Conflict of Interest**

The authors declare no conflict of interest.

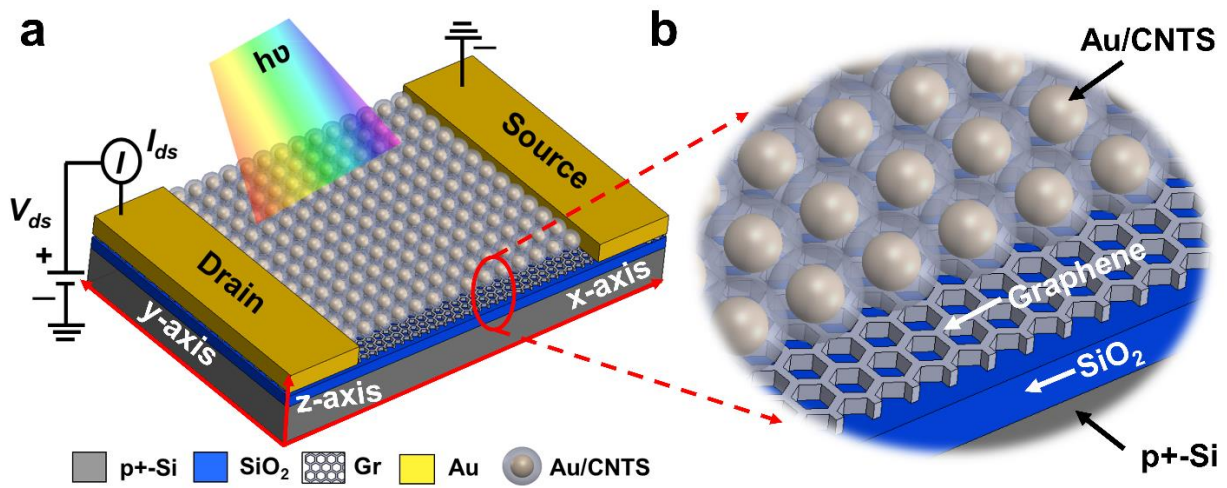

**Figure 1.** Schematic of Device: (a) Schematic illustration of Au/CNTS and graphene-based hybrid photodetector. (b) An enlarged view of the selected area of (a), showing core-shell Au/CNTS nanocrystals on the top of monolayer graphene. Here, Au/CNTS works as photoactive materials, and graphene works as a highly conductive channel layer.

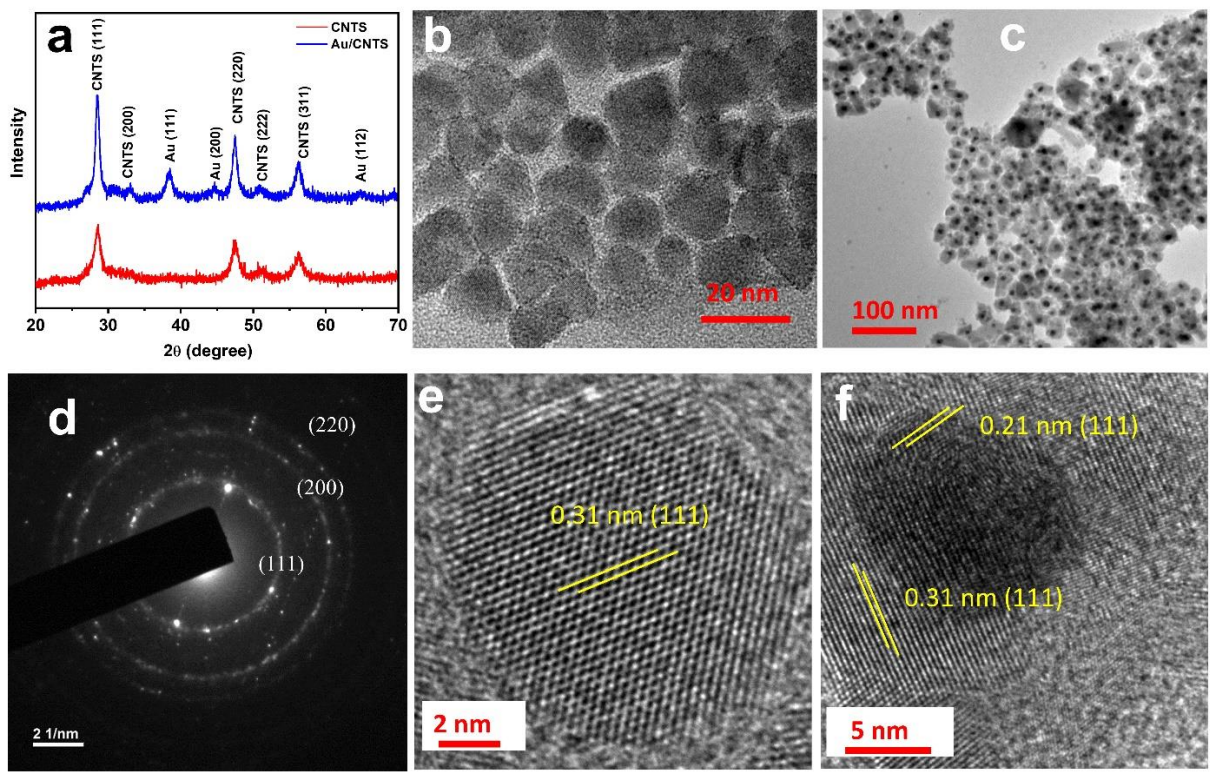

**Figure 2.** Structural characterization: (a) XRD patterns of CNTS and Au/CZTS NCs. (b) TEM images of CNTS NCs. (c) TEM images of Au/CNTS NCs with 0.05 mmol of Au. (d) Selected area electron diffraction (SAED) pattern of a single CNTS NCs. The concentric rings correspond to the significant peaks observed in the XRD pattern. (e) High-resolution TEM image of CNTS NCs. (f) High-resolution TEM image of Au/CNTS NCs.

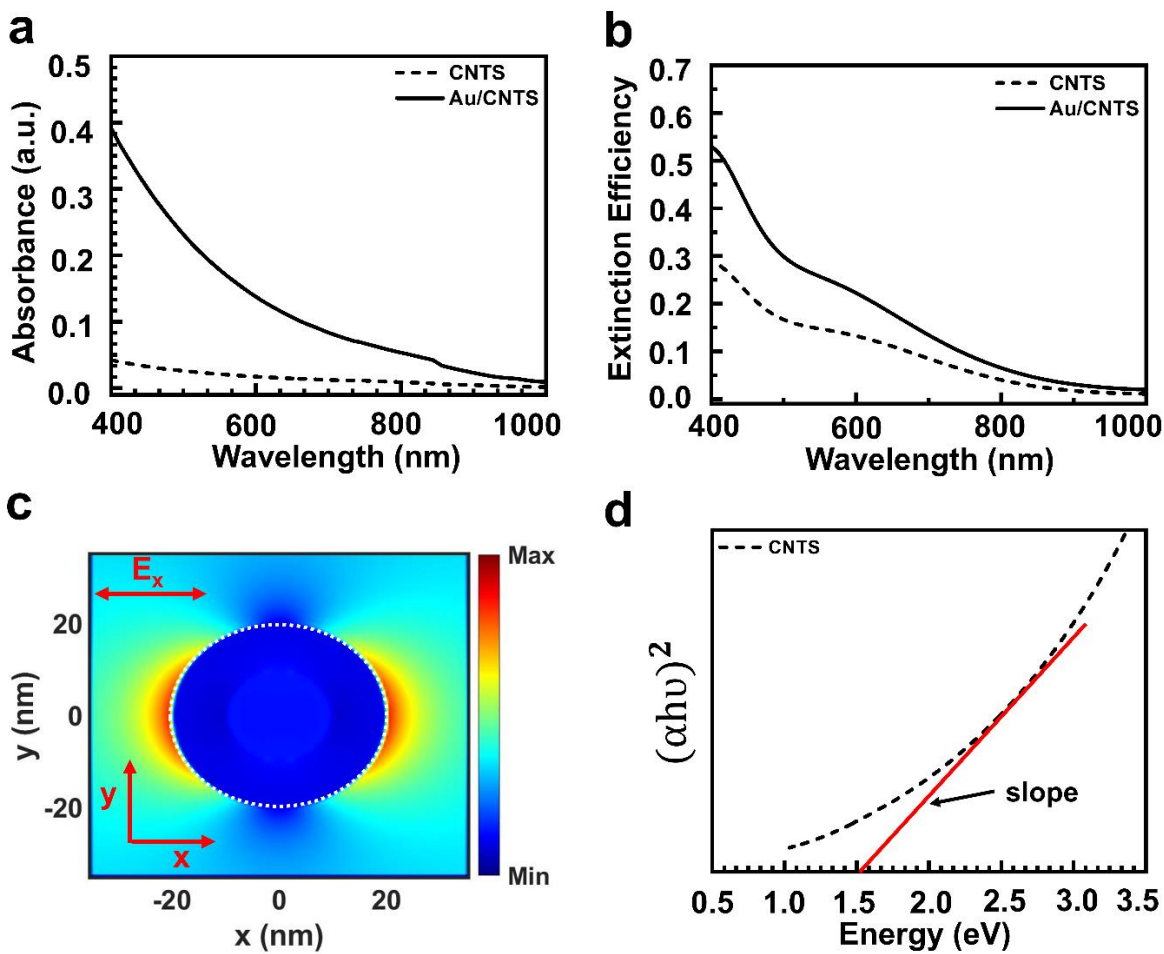

**Figure 3.** Optical characterization: (a) UV-Visible absorption spectra for CNTS and Au/CNTS. The optical absorption enhanced for Au/CNTS hybrid NCs. (b) Finite difference time domain (FDTD) simulated extinction efficiency for CNTS and Au/CNTS NCs. (c) Electric field distribution $E^2$ at resonance wavelength of 405 nm for Au/CNTS NCs indicating incident field localized around the NCs. (d) Tauc plot for band gap calculation of CNTS showing bandgap of 1.57 eV.

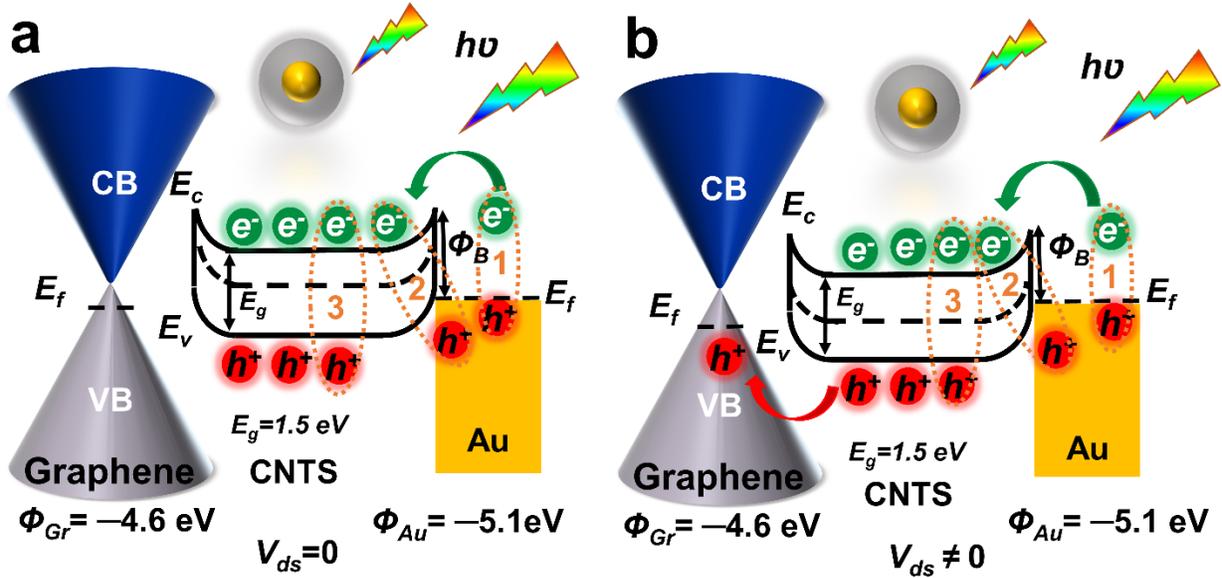

**Figure 4.** Energy band diagram of Au/CNTS NCs/graphene. (a) without applied bias $V_{ds} = 0V$ and (b) with applied bias ($V_{ds} > 0V$). The valance band, conduction band, and Fermi level of CNTS NCs and p-type graphene are shown by $E_v$, $E_c$, and $E_f$, respectively. The $\Phi_B$ represents the Schottky barrier height. The "h$^+$" and "e$^-$" symbols represent holes and electrons, respectively. The orange number 1,2,3 inside the dotted circle shows, the HET, PICTT, and PIRET effects, respectively. The green arrow represents the plasmonic-induced hot electrons transfer from the excited level of AuNPs. The red arrow indicates charge carrier transfer from CNTS NCs to the graphene. When $V_{ds} = 0$ V, there is no net current flow across the photodetector in contrast to the case for $V_{ds} > 0$ V.

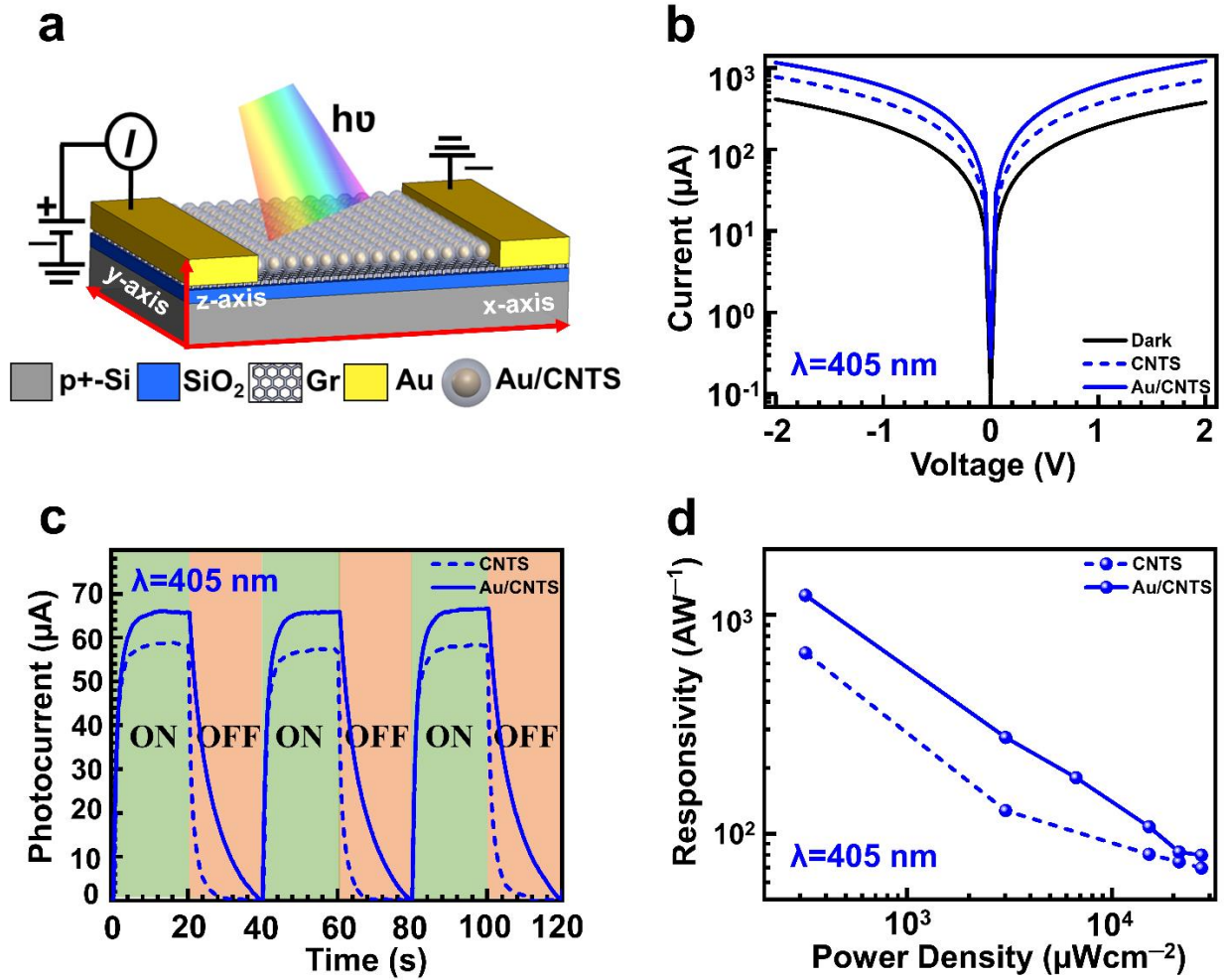

**Figure 5.** Optoelectronic characterization at 405 nm: (a) Schematic of the Au/CNTS/graphene-based hybrid photodetector. (b) Current with applied voltage characteristic of CNTS and Au/CNTS based hybrid photodetector at 860 µW incident power. (c) Transient photo-response of CNTS and Au/CNTS-based photodetector at a constant applied power density of 860 µW. (d) Responsivity for CNTS and Au/CNTS photodetector concerning incident power density.

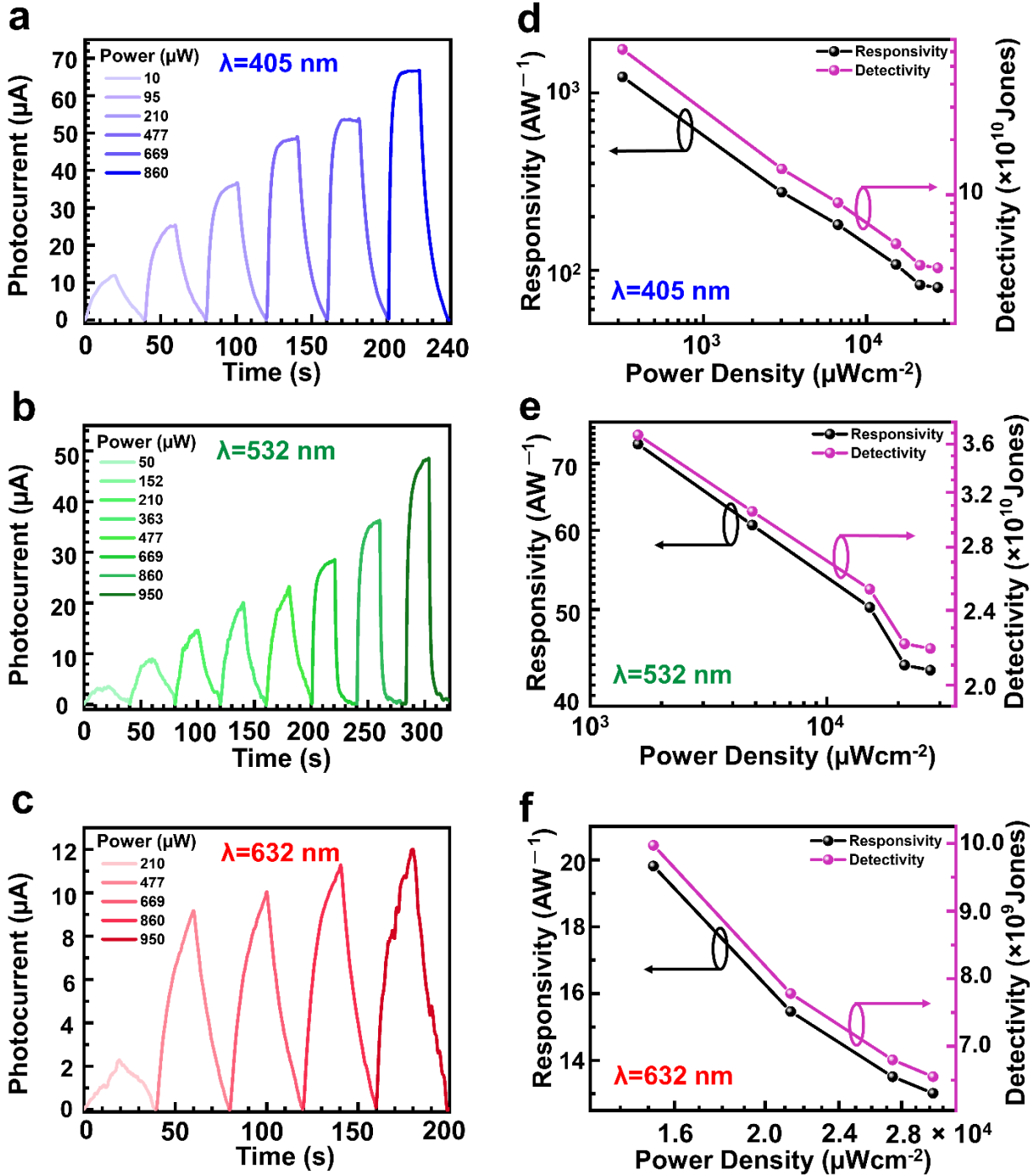

**Figure 6.** Optoelectronic characterization: (a-c) Transient photo-response for Au/CNTS-based hybrid photodetector with respect to the incident power at a wavelength of 405 nm, 532 nm, and 632 nm, respectively. (d-f) Responsivity (R) in the black axis and specific detectivity (D*) in the pink axis for Au/CNTS-based hybrid photodetector with incident power density at a wavelength of 405 nm, 532 nm, and 632 nm, respectively.

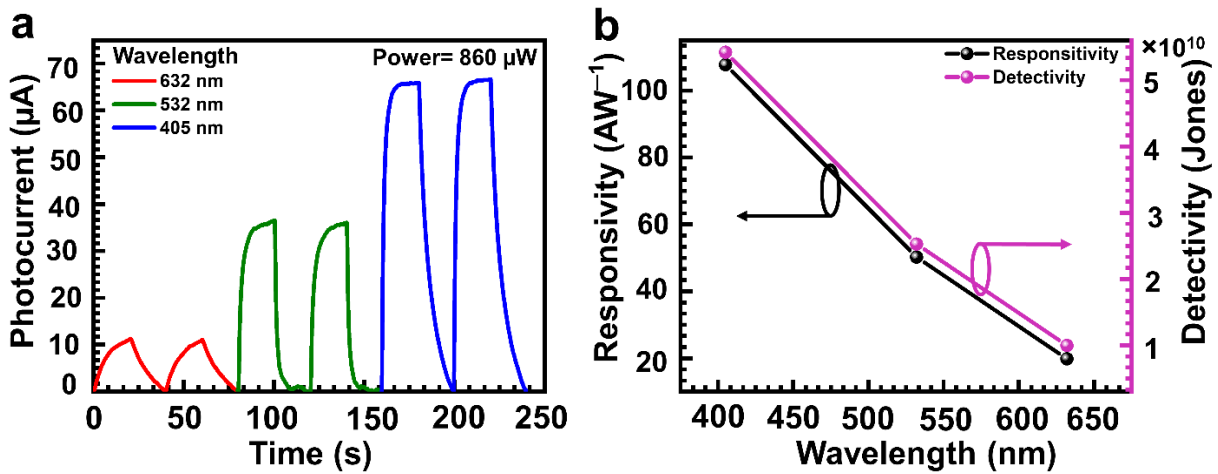

**Figure 7.** Transient photo-response for Au/CNTS-based hybrid photodetector with respect to power density for 405 nm, 532 nm, and 632 nm wavelength laser. (b) Wavelength-dependent responsivity (R) and specific detectivity (D*) (pink axis) at fixed power density for Au/CNTS-based hybrid photodetector at the incident power of 860 µW.